\documentclass[prl, twocolumn, showpacs, superscriptaddress]{revtex4}
\usepackage{amsmath,amsfonts, amsthm}
\usepackage{color}
\usepackage{bm}
\usepackage{mathrsfs}
\usepackage{graphicx, subfigure}
\usepackage{subfigure}
\usepackage{verbatim}
\usepackage{mathrsfs}
\usepackage{hyperref}

\begin{document}

\newcommand{\odiff}[2]{\frac{\di #1}{\di #2}}
\newcommand{\pdiff}[2]{\frac{\partial #1}{\partial #2}}
\newcommand{\di}{\mathrm{d}}
\newcommand{\ii}{\mathrm{i}}
\renewcommand{\vec}[1]{{\mathbf #1}}
\newcommand{\vx}{{\bm x}}
\newcommand{\ket}[1]{|#1\rangle}
\newcommand{\bra}[1]{\langle#1|}
\newcommand{\pd}[2]{\langle#1|#2\rangle}
\newcommand{\tpd}[3]{\langle#1|#2|#3\rangle}
\renewcommand{\vr}{{\vec{r}}}
\newcommand{\vk}{{\vec{k}}}
\renewcommand{\ol}[1]{\overline{#1}}
\newtheorem{theorem}{Theorem}
\newcommand{\mycomment}[1]{ {} }

\title{Topological Response Theory of Abelian Symmetry-Protected Topological Phases in Two Dimensions}
\author{Meng Cheng}
\affiliation{Station Q, Microsoft Research, Santa Barbara, CA 93106, USA}
\affiliation{Condensed Matter Theory Center, Department of Physics, University of Maryland, College Park, MD 20742}
\author{Zheng-Cheng Gu}
\affiliation{Institute for Quantum Information and Matter, California Institute
of Technology, Pasadena, CA 91125, USA}
\affiliation{Department of
Physics, California Institute of Technology, Pasadena, CA 91125,
USA}
\date{\today}
\begin{abstract}
  {It has been shown that} the symmetry-protected topological (SPT) phases with finite Abelian symmetries can be described by Chern-Simons field theory. We propose a topological response theory to uniquely identify the SPT orders, which allows us to obtain {a systematic scheme to classify} bosonic SPT phases with any finite Abelian symmetry group. We point out that even for finite Abelian symmetry, there exist bosonic SPT phases beyond the current Chern-Simons theory framework. We also apply the theory to fermionic SPT phases with $\mathbb{Z}_m$ symmetry and find the classification of SPT phases depends on the parity of $m$: for even $m$ there are $2m$ classes, $m$ out of which is intrinsically fermionic SPT phases and can not be realized in any bosonic system. 
  Finally we propose a classification scheme of fermionic SPT phases for any finite, Abelian symmetry.
\end{abstract}
\maketitle
{\it Introduction.}
In recent years, the research on topological matter has revealed a new class of gapped quantum phases, namely symmetry-protected topological(SPT) phases~\cite{Gu_PRB2009}. They are topologically distinct from trivial atomic insulators if (and only if) certain symmetries are not broken. A notable example is the electronic topological insulators in two and in three dimensions~\cite{Kane2005b, Bernevig2006, Fu_PRL07, Roy_PRB2009, Moore_PRB07, Kitaev2009}, {protected by time reversal and charge conservation symmetries}.  Although featureless in the bulk, SPT phases do support gapless boundary excitations protected by symmetry. This fact clearly distinguishes them from {a trivial product state or an atomic insulating state}. Recent theoretical discoveries, initiated by the group-cohomological construction~\cite{Chen_arxiv2011, Chen_science}, have vastly extended our knowledge on the classification of SPT phases. Systematic constructions of bosonic SPT phases, with arbitrary symmetry groups and in any spatial dimensions, have been proposed~\cite{Chen_PRB2011a, Fidkowski_PRB2011, Turner_PRB2011, Schuch_PRB2011, Chen_arxiv2011, Chen_science, Chen_PRB2011b}.  Later on, the construction was also generalized to fermionic systems, using the so-called group super-cohomology approach~\cite{Gu_arxiv2012}. In addition to the general classification, representative ground state wavefunctions, as well as exact solvable parent lattice Hamiltonians are also naturally derived. However, the wavefunctions and parent Hamiltonians are often quite complicated and it is not easy to access the low-energy universal properties of {these} phases, e.g., the edge properties.



Quite recently, field-theoretical approach have been taken to understand the physics of SPT phases in two dimensions protected by Abelian symmetries~\cite{Levin_arxiv2012,Lu_arxiv2012,  senthil_arxiv, senthil_3D, Liu_PRL2013}. There the SPT phases are described effectively by a multi-component $\mathbb{U}(1)$ Chern-Simons theory and the classification is derived from the equivalence classes of $\vec{K}$ matrix and the (perturbative) stability of edge theory. However, a (non-perturbative) bulk argument is missing and the equivalent classes of $\vec{K}$ matrix is also quite difficult to compute for an arbitrary Abelian symmetry group. On the other hand, the
underlying connection with (super)cohomology class is unclear as well.

In this work we present a unified treatment of both bosonic and fermionic SPT phases in two dimensions, with finite, unitary Abelian symmetries. Building upon the work of Levin and Gu~\cite{LevinGu_arxiv2012}, we formulate a topological response theory to probe the bulk properties of SPT phases.
This allows us to characterize
bosonic SPT phases with finite Abelian symmetry group that can be described in the Chern-Simons theory framework. We also study fermionic SPT phases with $\mathbb{Z}_m$ symmetry and interestingly find an even-odd effect: the classification is $\mathbb{Z}_m$ for odd $m$ and $\mathbb{Z}_{2m}$ for even $m$. Finally we obtain a minimal {set} of fermionic SPT phases for any finite, Abelian symmetry.

{\it Topological Response Theory of SPT Phases. } {We start with a brief overview of the underlying strategy for our classification scheme.
Because SPT phases are protected by a global symmetry, we can consider coupling the SPT phase to an external gauge field taking value in that symmetry group. Following the scheme proposed by Levin and Gu~\cite{LevinGu_arxiv2012},
{we use the braiding statistics of the gauge fluxes as a physical response to identify these SPT phases. }A classification can be achieved once the following two problems are resolved: (a) How to compute the braiding statistics of a gauge flux. (b) How to identify the equivalence classes of the flux statistics. Heuristically, when inserting a flux into a SPT phase, the statistics of the flux depends on the charged particles bind to the flux due to unknown local energetics. Therefore, each SPT phase should be associated with a whole family of flux statistics obtained from attachment of gauge charges, i.e. the whole flux sector.} 

Now we describe the first part of our proposal, i.e. coupling the SPT phase to an external gauge field. We review the Chern-Simons field theory of SPT phases~\cite{Lu_arxiv2012, Levin_arxiv2012}. They are described by a multi-component Abelian Chern-Simons theory, which in its most general form is given by the following Lagrangian:
\begin{equation}
  \mathcal{L}_\text{CS}=\frac{1}{4\pi}\varepsilon^{\mu\nu\lambda}a_\mu^I K_{IJ}\partial_\nu a_\lambda^J.
  \label{}
\end{equation}
Here $\vec{K}$ is a $N\times N$ integer symmetric matrix. Notice that for bosonic systems, all diagonal entries of $\vec{K}$ must be even while at least one of the diagonal entries is odd for fermionic systems. Quantization of this gauge theory gives $|\det \vec{K}|$ ground states on a torus. Since a SPT phase can not have any intrinsic topological order, we require $\det \vec{K}=\pm 1$ so there is no topological degeneracy on torus. And $\vec{K}$ should have equal number of positive and negative eigenvalues to avoid chirality.

We will now assume that the system under consideration has a global, on-site symmetry group $G$. We consider unitary, finite Abelian group, which generally can be written as $G=\mathbb{Z}_{m_1}\times\mathbb{Z}_{m_2}\times\cdots\mathbb{Z}_{m_k}$ where $m_k>1$. We assume that the $\mathbb{Z}_{m_j}$ subgroup is generated by $g_j$ with $ g_j^{m_j}=1$.  The matter fields carry irreducible representations of the symmetry group. Since the symmetry is Abelian and finite, it amounts to assign $\mathbb{Z}_{m_1}\times\mathbb{Z}_{m_2}\times\cdots\mathbb{Z}_{m_k}$ charges $q_\alpha^I$ to the $I$-th matter field. Here $q^I_\alpha$are  valued in $\{0,1,\dots, m_\alpha-1\}$ and the subscript $\alpha$ refers to symmetry subgroups. Different assignment of the charges (or equivalently, symmetry transformation properties) can lead to distinct symmetry-protected phases. 
{Next we} couple the SPT phase to a gauge field taking value in the gauge group $G$. This is achieved by writing down minimal coupling for the matter fields in the SPT phases, uniquely determined by the charges carried by the matter fields.

To facilitate the continuum field theory formulation, we view the group $G$ as a discrete subgroup of $\mathbb{U}(1)^k$. This perspective allows us to introduce $k$ external $\mathbb{U}(1)$ gauge fields $A^\alpha_\mu,\alpha=1,\dots,k$ that minimally couple to the matter fields in the SPT phases and then by introducing $k$ Higgs fields $\varphi_\alpha$ with charge $m_\alpha$ we can obtain a continuum version of discrete gauge theory as the symmetry broken phase of the $\mathbb{U}(1)^k$ Higgs theory. The presence of Higgs condensation results in quantization of gauge fluxes.

After we gauge the {global} symmetry, the minimal coupling between the matter fields in the bulk and the external gauge fields reads:
\begin{equation}
  \mathcal{L}_\text{coupling}=\sum_\alpha\vec{q}_{\alpha}^T\vec{j}{A}^\alpha
  \label{}
\end{equation}
The full theory is given by $\mathcal{L}=\mathcal{L}_\text{CS}+\mathcal{L}_\text{coupling}$. We then substitute $\vec{j}^\mu=\frac{1}{2\pi}\varepsilon^{\mu\nu\lambda}\partial_\nu \vec{a}_\lambda$ and integrate out internal gauge fields $\vec{a}$, yielding an effective theory for the external fields $A^\alpha$:
\begin{equation}
  \mathcal{L}_\text{eff}=\frac{1}{4\pi}\varepsilon^{\mu\nu\lambda}{A}^\alpha_\mu\vec{q}_\alpha^T\vec{K}^{-1}\vec{q}_\beta\partial_\nu {A}_\lambda^\beta+\mathcal{L}_\text{Higgs}[\varphi_\alpha, A_\alpha].
  \label{eqn:gaugedL}
\end{equation}
We denote $\tilde{\vec{K}}_{\alpha\beta}=\vec{q}_\alpha^T\vec{K}^{-1}\vec{q}_\beta$ in the following discussion.
We have therefore derived an Abelian Chern-Simons-Higgs theory as the effective ``response theory'' for the gauged SPT phases. Physically, as long as the symmetry $G$ remains unbroken, we can always gauge the SPT phases. The braiding statistics of the fluxes are fully captured by the effective theory \eqref{eqn:gaugedL}. 

Let us now identify the equivalence classes of the flux statistics, as advertised in the beginning of our presentation.  We start from the simplest case $G=\mathbb{Z}_m$. Then we just have a single Chern-Simons term at the level $\tilde{K}=\vec{q}^T\vec{K}^{-1}\vec{q}$.  An ``elementary'' vortex in the Higgs field encloses $\frac{2\pi}{m}$ gauge flux, as a result of flux quantization.
The exchange statistics of the vortices can be calculated (see the Supplementary Material~\cite{suppl} for details)
:  $\theta=-\frac{\pi \tilde{K}}{m^2}$.
To claim $\theta$ (or equivalently, $\tilde{K}$) as a unique ``topological invariant'' of the $\mathbb{Z}_m$ SPT phases, we must understand when two seemingly different values of  $\theta$ actually describe the same topological class. To this end, notice that $\theta$ is the statistical angle of a pure Higgs vortex. Depending on local energetics, there may be $\mathbb{Z}_m$ charges bound to it.  Assuming  $q$ elementary $\mathbb{Z}_m$ charges are attached to the vortex, the composite object has a statistical angle $\theta+\frac{2\pi q}{m}$.
This implies that the exchange statistics of a $\mathbb{Z}_m$ flux as a ``topological invariant'' for SPT phases is defined modulo $\frac{2\pi}{m}$, or $\tilde{K}$ is defined modulo $2m$. Two SPT phases whose corresponding statistical angles differ by an integer multiple of $\frac{2\pi}{m}$ should be considered as being equivalent, since the statistics can be changed by just adjusting the local energetics without affecting the bulk. Therefore the equivalance classes of $\tilde{K}$ is at most $\mathbb{Z}_{2m}$.


However, this is not the end of the story. The statistics of the underlying particles, bosons or fermions, make a big difference. We first consider the simpler case of bosonic systems.
One can easily see that $\tilde{K}$ for bosons must be even.
Thus the classification is reduced to $\mathbb{Z}_m$.
They can be realized by the following $\{\vec{K},\vec{q}_g\}$:
{\begin{equation}
  \vec{K}=
  \begin{pmatrix}
	0 & 1\\
	1 & -2n
  \end{pmatrix}, \vec{q}_g=
  \begin{pmatrix}
	1\\
	0
  \end{pmatrix}, n=0,1, \dots, m-1.
  \label{eqn:znspt}
\end{equation}
It corresponds to $\theta=\frac{2\pi n}{m^2}$.}  We notice that if the external gauge field is treated dynamically, the intrinsic topological order described by \eqref{eqn:gaugedL} is essentially a twisted {gauge theory} with gauge group $G$ ~\cite{Propitius_thesis,twistedQD} (see the Supplementary Material~\cite{suppl} for the derivation of this fact). {However, it is important to emphasize that the classification of SPT phases using the braiding statistics of the gauge fluxes is {\it not} equivalent to classifying the intrinsic topological orders described by the corresponding Chern-Simons-Higgs theory Eq. \eqref{eqn:gaugedL} (also see the Supplementary Material~\cite{suppl} for more discussion).}

\mycomment{
\textbf{[Meng: A dynamic gauge theory without matter fields has a non-local Hilbert space and is not equivalent to the corresponding twisted quantum double models with intrinsic topological order and emergent gauge fields from a standard perspective. I think if we treat the external guage theory to be dynamic, we just derive the Dijkgraaf-Witten gauge theory for any finite group. The Dijkgraaf-Witten gauge theory and SPT order is one to one correspondence.
Maybe this disagreement is because we use a Higgs field here instead of a discrete dynamic guage field and lead to Eq.(6) in the supplementary material.
 However, I feel the discussion here might be confusing here anyway and it is impossible to clarify it without a careful explanation.
 May be we just remove this part, since it is not necessary for this paper and will confuse readers.Or you can just put it in the appendix if you really want to clarify this point, but it is not suitable to call a dynamical guage theory as intrinsic topological order.]}
To classify the intrinsic topological order of the emergent gauge theory, only the braiding statistics is concerned and we do not keep track of the charge and flux carried by the quasiparticles at all. While in the response theory, we are not allowed to arbitrarily change the flux sectors.
The difference is already manifested in the case $G=\mathbb{Z}_m$. For example, when $m=5$ two distinct SPT phases, $n=2$ and $n=3$ in Eq. \eqref{eqn:znspt}  lead to topological gauge theories with identical intrinsic topological orders. In fact this phenomena occurs for all odd $m$ and some even $m$ (see the Supplementary Material~\cite{suppl} for more details).
This issue has not been clarified in some related works~\cite{Ying_2012,Hung_2012} in the context of classifications of symmetry-enriched topological orders, however, it is manifested in our approach.}

Having understood $G=\mathbb{Z}_m$, we move to the next level of complexity $G=\mathbb{Z}_m\times\mathbb{Z}_n$. In this case, we have to consider  $\mathbb{U}(1)\times \mathbb{U}(1)$ external gauge fields and the response theory is fully characterized by a $2\times 2$ matrix $\tilde{\vec{K}}$. Repeating our argument above, the two diagonal elements determine the exchange statistics of the two types of gauge fluxes corresponding to the two subgroups and give $\mathbb{Z}_m\times\mathbb{Z}_n$ classification. The off-diagonal element introduces a new ingredient, the braiding statistics between the two gauge fluxes, denoted by $\theta_{12}$:
  $\theta_{12}=\frac{2\pi \tilde{K}_{12}}{mn}$.
To determine the equivalence classes of $\theta_{12}$, we notice that we can attach $\mathbb{Z}_m$ charges to a $\mathbb{Z}_n$ flux or vice versa and the mutual statistical angles are changed by integer multiples of $\frac{2\pi}{m}$ and $\frac{2\pi}{n}$, respectively. Therefore we identify the equivalence relation as
\begin{equation}
  \theta_{12}\equiv \theta_{12}+\frac{2\pi k_1}{m}+\frac{2\pi k_2}{n}, k_1,k_2\in\mathbb{Z}.
  \label{}
\end{equation}
In terms of $\tilde{K}_{12}$,
\begin{equation}
  \tilde{K}_{12}\equiv \tilde{K}_{12}+k_1n+k_2m.
  \label{}
\end{equation}
{The set of integers generated by $k_1 n+ k_2 m, k_1, k_2\in\mathbb{Z}$ is simply the integer multiples of $(m,n)$}. Therefore we conclude that that $\tilde{K}_{12}$ is defined modulo $(m,n)$, which gives additional $\mathbb{Z}_{(m,n)}$ classes. The complete classification is thus $\mathbb{Z}_m\times\mathbb{Z}_n\times\mathbb{Z}_{(m,n)}$.

What is the physical meaning of the mutual braiding statistics in the SPT phases? To give a concrete example, consider a SPT phase characterized by $\tilde{K}_{11}=\tilde{K}_{22}=0, \tilde{K}_{12}=l$. One can easily write down the following Chern-Simons theory:
\begin{equation}
  \vec{K}=
  \begin{pmatrix}
	 0 & 1\\
	 1 & 0
  \end{pmatrix},
\vec{q}_{1}=
\begin{pmatrix}
  0\\ 1
\end{pmatrix},
 \vec{q}_{2}=
\begin{pmatrix}
  l\\ 0
\end{pmatrix}
  \label{eqn:znzn}
\end{equation}
So the two ``dual'' degrees of freedoms in the SPT phases carry the two subgroup symmetries respectively. Other classes can be understood in a similar way.
In general, all bosonic SPT phases with $\mathbb{Z}_m\times \mathbb{Z}_n$ symmetry are realized by~\cite{note1}
\begin{equation}
  \vec{K}=\sigma_x\otimes\mathbf{1}_{2\times 2}, \vec{q}_1=
  \begin{pmatrix}
	0\\ 1\\ 1\\ p
  \end{pmatrix},
  \vec{q}_2=
  \begin{pmatrix}
	l \\ 0 \\ 1\\ q
  \end{pmatrix}.
  \label{}
\end{equation}
It corresponds to $\tilde{\vec{K}}=\begin{pmatrix} 2p & l+p+q\\ l+p+q & 2q\end{pmatrix}$.

  We are now well prepared to generalize the above picture to arbitrary finite Abelian group $G$ with $k\geq 2$ generators. The $\tilde{\vec{K}}$ matrix of the response theory has $k$ diagonal elements, yielding $\mathbb{Z}_{m_1}\times\mathbb{Z}_{m_2}\times\cdots\mathbb{Z}_{m_k}$ classification. The $\frac{k(k-1)}{2}$ off-diagonal elements determine the braiding statistics between the gauge fluxes and the classification is $\prod_{i<j}\mathbb{Z}_{(m_i,m_j)}$. Therefore, we conclude that for a given symmetry group $G=\prod_{i=1}^k \mathbb{Z}_{m_i}$, the Abelian Chern-Simons theory construction can give $\prod_{i=1}^k \mathbb{Z}_{m_i}\times\prod_{ i<j} \mathbb{Z}_{(m_i,m_j)}$ classification of possible SPT phases. In fact, what we have obtained turns out to be a subset of the cohomological classification~\cite{Chen_arxiv2011}, which gives additional $\prod_{i<j<k}\mathbb{Z}_{(m_i, m_j, m_k)}$ classes
\footnote{For ${G}=\prod_{i=1}^k \mathbb{Z}_{m_i}$, one finds~\cite{Propitius_thesis}
 $H^3(G, \mathbb{U}(1))=\prod_{i} \mathbb{Z}_{m_i}\prod_{i<j} \mathbb{Z}_{(m_i,m_j)}\prod_{i<j<k}\mathbb{Z}_{(m_i, m_j, m_k)}$.
Clearly, when $k\geq 3$ the Abelian Chern-Simons theory approach fails to capture the classes of SPT phases associated with $\mathbb{Z}_{(m_i, m_j, m_k)}$. Physically, this is because the gauge fluxes in the topological response theory carry non-Abelian statistics~\cite{Propitius_thesis}.}.

{It is now clear what is the limitation of the present approach: Since the currents in the SPT phase all carry Abelian charges, the gauged theory is bound to be Abelian. A simple example that goes beyond this scheme is $G=\mathbb{Z}_m\times H$ with $H$ having appropariate projective representations (e.g $H=\mathbb{Z}_m\times\mathbb{Z}_m$). Then the gauge flux of the $\mathbb{Z}_m$ subgroup can transform projectively under the other subgroup $H$, which implies symmetry-protected degeneracy associated with the flux~\cite{Chen_2013}. Consequently the gauged theory exhibits certain non-Abelian character. 
We leave the study of these phases for future publications.
}


{\it Fermionic SPT Phases.}
We now study fermionic SPT phases protected by a symmetry group $G=\mathbb{Z}_m$. 
We will show quite interestingly, the classification exhibits an even-odd effect: for odd $m$ the classification is $\mathbb{Z}_m$ while for even $m$, it is $\mathbb{Z}_{2m}$. 

The main difference between fermionic systems and bosonic ones is that the ``fundamental'' particles are fermions which themselves have a statistical angle $\pi$. { Therefore, regardless of the SPT structure, the statistical angle of a flux can always be changed by $\pi$ via attachment of an odd number of fundamental fermions~\cite{Gu_2010ftop, Freedman_AP2004}. This has profound consequence on the the equivalence classes of the braiding statistics of fluxes in the gauged theory. In fact, we already see there is a difference between even and odd $m$, since in a $\mathbb{Z}_m$ gauge theory with even $m$, it is always possible to change the statistical angle of fluxes by $\pi$ via attaching certain number of gauge charges (i.e. there are dyonic excitations with fermionic statistics) while for odd $m$ this is not the case.}

We first consider $m$ odd. Attachment of (bosonic) $\mathbb{Z}_m$ charges can change the statistical angle $\theta$ of an elementary gauge flux by $\frac{2\pi p}{m}, p\in \mathbb{Z}$. {Further allowing possible attachment of $\mathbb{Z}_m$-neutral fermions, $\theta$ can be changed by $ \frac{\pi p}{m}$. } One might wonder what if all fundamental fermions are $\mathbb{Z}_m$-charged. In this case, we can attach a composite of $m$ such fermions, which is $\mathbb{Z}_m$-neutral and still has fermionic statistics since $m$ is odd. We therefore need to identify $\theta$ with $\theta+\frac{\pi p}{m}$. As a result, given two SPT phases characterized by $\tilde{K}_1$ and $\tilde{K}_2$, the equivalence relation now becomes:
\begin{equation}
  \frac{\pi\tilde{K}_1}{m^2}-\frac{\pi\tilde{K}_2}{m^2}=\frac{\pi p}{m},
  \label{}
\end{equation}
which implies that $\tilde{K}\equiv \tilde{K}+{mp}$, in constrat to $\tilde{K}\equiv \tilde{K}+{2mp}$ for bosonic SPT phases. We then easily see that there are only $m$ different classes ({Notice we allow both even and odd $\tilde{K}$ in the fermionic case}).
 In addition, all fermionic SPT phases with $\mathbb{Z}_m$ symmetry can be identified with the bosonic SPTs together with trivial fermions (see the Supplementary Material for the derivation~\cite{suppl}).
 {A physical interpretation is the following: bosonic SPT phases can be realized in any fermionic system when strong on-site interactions completely suppress charge fluctuations and effectively we have a spin system.} 

 Now we turn to the case when $m$ is even.  As we have elaborated, attachment of $\mathbb{Z}_{m}$ charges allows the statistical angle of fluxes to be changed by $\frac{2\pi p}{m}$. The fact that we now have fundamental fermions does not introduce additional equivalence conditions, {since for $p=m/2$ we already have a fermionic charge-flux composite.} The equivalence relation stays the same as the one for bosonic SPT phases, but now we allow both even and odd $\tilde{K}$ so there are $2m$ distinct $\tilde{K}$. When $\tilde{K}$ is odd, the corresponding SPT phases are intrinsically fermionic and {can not be realized in any bosonic system}. We are then led to the conclusion that the classification for $m$ even is $\mathbb{Z}_{2m}$.
We notice that the $m=2$ case receives a lot of attention recently~\cite{Qi_arxiv2012, Ryu_arxiv2012, Yao_arxiv2012,Gu_2013} since it serves as a nice example of ``collapse'' of classification of non-interacting fermions (which is $\mathbb{Z}$ in the present case) when interactions are taken into account ($\mathbb{Z}_8$). The Chern-Simons field theory approach {gives rise to a} $\mathbb{Z}_4$ classification. The missing four classes are those with unpaired Majorana edge modes which are beyond the Chern-Simons field theory description.

Having worked out the classification for $G=\mathbb{Z}_m$, it is straightforward to generalize to arbitrary finite Abelian group $G=\prod_\alpha \mathbb{Z}_{m_\alpha}$. We notice that the underlying fermionic statistics does not effect mutual braiding properties, so the additional classes due to nontrivial mutual braiding statistics between gauge fluxes in different conjugate classes are still classified by $\prod_{i<j}\mathbb{Z}_{(m_i,m_j)}$. Thus we obtained a classification of fermionic SPT based on Chern-Simons theory as $\prod_i \mathbb{Z}_{m_i^*}\prod_{i<j}\mathbb{Z}_{(m_i, m_j)}$ where $m^*$ is defined as $m^*=m$ for odd $m$ and $2m$ for even $m$.

{\it Edge theory for fermionic SPT phase.}
Having established the classification of fermionic SPT phases with $\mathbb{Z}_m$ symmetry,  we move on to study the edge theory and its stability in more detail.
Let us consider the simplest one of $\mathbb{Z}_m$(for even $m$) fermionic SPT phases described by the following $\vec{K}$ matrix and $\vec{q}$ vector: 
\begin{equation}
  \vec{K}=
  \begin{pmatrix}
	 1 & 0\\
	 0 & -1
  \end{pmatrix},
\vec{q}=
\begin{pmatrix}
  1\\ 0
\end{pmatrix},
\label{eqn:fspt}
\end{equation}

It is well known that the Chern-Simons theory implies existence of gapless edge states, whose effective Lagrangian can be derived from gauge invariance principle~\cite{Wen_AdvPhys1995}:
\begin{equation}
  \mathcal{L}_\text{edge}=\frac{1}{4\pi}(\partial_t\phi_I K_{IJ}\partial_x \phi_J-\partial_x\phi_I V_{IJ}\partial_x \phi_J).
  \label{}
\end{equation}
with symmetry transformation:
$\bm{\phi} \rightarrow \bm{\phi}+\frac{2\pi}{m}\vec{q}$.
To include interaction effect, it is convinient to switch to a non-chiral basis $\phi_{1}= \varphi-\theta, \phi_{2}=\varphi+\theta$. We then have a generic Luttinger liquid model of the gapless edge:
\begin{equation}
  H=\int\di x\,\frac{u}{2\pi}\left[ K(\partial_x\theta)^2+K^{-1}(\partial_x\varphi)^2 \right].
  \label{}
\end{equation}
Here $u$ is the charge velocity and $K$ the Luttinger parameter.
One can see that the $\mathbb{Z}_m$ transformation acts on the non-chiral bosonic fields as $\varphi\rightarrow\varphi+\frac{\pi}{m}, \theta\rightarrow\theta+\frac{\pi}{m}$. To understand the stability of the edge theory, we add the leading perturbations $\cos 2m\varphi$ and $\cos 2m\theta$ allowed by $\mathbb{Z}_m$ symmetry. They have scaling dimensions $\frac{m^2K}{2}$ and $\frac{m^2}{2K}$respectively. So demanding that these two perturbations are irrelevant, we find the stable region is $\frac{2}{m^2}<K<\frac{m^2}{2}$. Interestingly, if we create domain walls on the edge such that the two sides have distinct $\mathbb{Z}_m$-breaking mass gaps (i.e. $\varphi$ condensed or $\theta$ condensed), a localized Majorana zero mode has to appear on the domain wall in those intrinsically fermionic SPT phases, which can serve as an experimental signature.

{\it Conclusion and Discussion.}
In conclusion, we systematically investigate SPT phases with an Abelian finite group symmetry within the framework of Chern-Simons field theory. We develop a topological response theory to classify the SPT phases by gauging the symmetry group. A careful examination of the equivalence classes of the braiding statistics of gauge fluxes enables us to characterize all possible bosonic SPT phases that can be realized as Abelian Chern-Simons theories.
We also compare our approach with the results of the group cohomology theory and discuss the limitation of $K$-matrix construction. Indeed, the topological response theory describes a non-perturbative effect in the bulk of SPT phases and provides us a unique way to identify different SPT phases. Finally, we extend the classification scheme to fermionic SPT phases. For the simplest symmetry group $G=\mathbb{Z}_m$, we find the classification of fermionic SPT phases has an intriguing even-odd dependence on $m$. We then generalize the classification to arbitrary finite Abelian groups. We also discuss the edge stability of those intrinsic fermionic SPT phases. For future studies, it would be very interesting to develop a topological response theory to describe those SPT phases with non-Abelian flux statistics, as well as possible extension to anti-unitary symmetries. On the other hand, the concept of topological response theory is also very useful for the classification of symmetry enriched topological(SET) order.

{\it Acknowledgement.} MC thanks Lukasz Fidkowski, Chetan Nayak, Zhenghan Wang and Juven Wang for insightful discussions.  ZCG is supported in part by Frontiers Center with support from the Gordon and Betty Moore Foundation.

{\it Note added.} By the completion of this work, we became aware of recent preprints~\cite{Lu_arxiv2013, Wen2013} which have some overlap with our results.


\begin{thebibliography}{34}
\bibitem[{not()}]{note1}
\bibinfo{note}{We note by pass that $2\times 2$ $\vec{K}$ matrix is not
  sufficient to realize all the SPT phases in this case if $(m,n)>1$. For
  example, when $m=n=2$, $7$ out of the $8$ SPT phases can be realized by
  $2\times 2$ $\vec{K}$ matrix but the one with
  $\tilde{K}_{11}=\tilde{K}_{22}=2,\tilde{K}_{12}=1$ needs minimally $4\times
  4$ $\vec{K}$ matrix.}
\end{thebibliography}

\newpage

\onecolumngrid

\vspace{1cm}
\begin{center}
{\bf\Large Supplementary material}
\end{center}
\vspace{0.2cm}


\renewcommand{\theequation}{S\arabic{equation}}
\setcounter{equation}{0}

\section{Abelian Chern-Simons-Higgs Theory}
In the main text we have argued that the gauged SPT phases can be described by an Abelian Chern-Simons theory with the $\mathbb{U}(1)$ gauge fields Higgsed. In this section we provide derivations of the flux statistics in the Abelian Chern-Simons-Higgs theory.

 Recall that after we integrate out the gapped degrees of freedom in the SPT phases protected by a symmetry group $G=\prod_{\alpha=1}^k \mathbb{Z}_{m_\alpha}$, we arrive at the following Lagrangian density:
\begin{equation}
  \mathcal{L}_{\text{CSH}}=\frac{1}{4\pi}\varepsilon^{\mu\nu\lambda}A^\alpha_\mu \tilde{K}_{\alpha\beta}\partial_\nu A^\beta_\lambda-\sum_{\alpha}\Big[\frac{1}{2} \big|(\partial_\mu-im_{\alpha}A_\alpha) \varphi_\alpha)\big|^2-V(\varphi_\alpha)\Big].
  \label{}
\end{equation}
Here $A_\alpha$ are external $\mathbb{U}(1)^k$ gauge fields, which we take to be semi-classical, adiabatically varying background fields.

We assume that the amplitudes of the Higgs fields are fixed by the potential energy term $V(\varphi_\alpha)$ and only the phase degrees of freedom remain. First we take this energy scale to be infinite. Heuristically speaking, the gauging is in the ``weak'' sense and the fluxes are semi-classical objects. Then $A$ field does not have its own dynamics. In fact, it should be thought as being slaved to the vortex current $j_{v,\alpha}$. In other words, we can integrate over $A$ but with a contraint $\frac{1}{\pi}\varepsilon^{\mu\nu\lambda}\partial_\nu A_\lambda=j_v^\mu$ enforced. With this in mind we can concentrate on the Chern-Simons part and introduce Lagrange multipliers $a_\mu$ to resolve the constraint:
\begin{equation}
	\mathcal{L}_\text{eff}=
	\frac{1}{4\pi}\varepsilon^{\mu\nu\lambda}A^\alpha_\mu \tilde{K}_{\alpha\beta}\partial_\nu A^\beta_\lambda+a^\alpha_{\mu}\left(\frac{m_\alpha}{\pi}\varepsilon^{\mu\nu\lambda}\partial_\nu A_{\alpha\lambda}-j_{v,\alpha}^\mu\right).
	\label{}
\end{equation}
Now we can integrate out $A_\mu$ yielding a Chern-Simons action for the gauge fields $a_\mu$:
\begin{equation}
	\mathcal{L}_\text{eff}=-\frac{m_\alpha {\tilde{K}}^{-1}_{\alpha\beta}m_\beta}{4\pi}\varepsilon^{\mu\nu\lambda}a^\alpha_\mu \partial_\nu a^\beta_\lambda-a^\alpha_\mu j_{v,\alpha}^\mu.
	\label{}
\end{equation}
We can then compute the braiding statistics of the fluxes directly from the action.

Now we step back and derive the full dynamical gauge theory. We write $\varphi_\alpha=v_\alpha e^{i\theta_\alpha}$ and substitute into the Lagrangian density:
\begin{equation}
  \mathcal{L}_\text{eff}=\frac{1}{4\pi}\varepsilon^{\mu\nu\lambda}A^\alpha_\mu \tilde{K}_{\alpha\beta}\partial_\nu A^\beta_\lambda-\sum_{\alpha}\frac{v_\alpha^2}{2}\Big(m_{\alpha} A_\mu^\alpha-{\partial_\mu\theta_\alpha}\Big)^2.
  \label{}
\end{equation}
First we perform the Hubbard-Stratonovich transformation of the quadratic term $\propto (mA-\partial\theta)^2$ and write
\begin{equation}
   \mathcal{L}_\text{eff}=\frac{1}{4\pi}\varepsilon^{\mu\nu\lambda}A^\alpha_\mu \tilde{K}_{\alpha\beta}\partial_\nu A^\beta_\lambda-\sum_\alpha\Big[\frac{1}{v_\alpha^2}\xi_\alpha^2-\xi_\alpha^\mu\Big(m_{\alpha} A_\mu^\alpha-{\partial_\mu\theta_\alpha}\Big)\Big].
  \label{}
\end{equation}
Here $\xi_\mu$ is the Hubbard-Stratonovich field. Then decompose the phase field as $\theta_\alpha=\eta_\alpha+\zeta_\alpha$ where $\eta_\alpha$ is the smooth part of the phase fluctuation and $\zeta_\alpha$ is the singular(vortex) part determined by $j_v$. Integrate out the smooth part of the phase fields $\eta_\alpha$, we obtain the constraint $\partial_\mu\xi_\alpha^\mu=0$, which can be resolved as $\xi_\alpha^\mu=\frac{1}{2\pi}\varepsilon^{\mu\nu\lambda}\partial_\nu b_{\alpha\lambda}$.
So we obtain the following dual representation
\begin{equation}
	\mathcal{L}=\frac{1}{4\pi}\varepsilon^{\mu\nu\lambda}A^\alpha_\mu \tilde{K}_{\alpha\beta}\partial_\nu A^\beta_\lambda+\frac{m_{\alpha}}{2\pi}\varepsilon^{\mu\nu\lambda}A^\alpha_{\mu}\partial_\nu b^\alpha_{\lambda}+\frac{1}{2\pi}\varepsilon^{\mu\nu\lambda}j_\mu\partial_\nu A_{\alpha\lambda}+\varepsilon^{\mu\nu\lambda}j_{v,\mu}\partial_\nu b_{\alpha\lambda}-\sum_\alpha\frac{1}{4\pi^2v_\alpha^2}(\partial_\mu b_{\alpha\nu}-\partial_\nu b_{\alpha\mu})^2.
  \label{}
\end{equation}
Here we have also included the charge current $j$ coupled to $A$. The Maxwell term is less relevant than the Chern-Simons type terms and can be safely neglected.
The remaining action is a doubled Chern-Simons theory. We can  pack the theory into a $K$-matrix:
\begin{equation}
	K=
	\begin{pmatrix}
		0 & \vec{m}\\
		\vec{m} & \tilde{\mathbf{K}}
	\end{pmatrix}.
	\label{eqn:kmat}
\end{equation}
Here $\vec{m}=[m_1, m_2, \dots]$ is a diagonal matrix. The statistics of the charge and flux excitations can be easily computed by taking the inverse of $K$. This also establishes formally the connection between the SPT phases and the Abelian intrinsic topological order given by the $K$ matrix given in \eqref{eqn:kmat}.

\section{Intrinsic Topological Order and Classification of Topological Responses}
We elaborate on the nonequivalence between the classification of the gauged SPT phases as a ``topological response'' theory and the classification of the intrinsic topological order.  As emphasized in the main text,  the classification of the response theory is not completely equivalent to the classification of the intrinsic topological order defined by the Chern-Simons-Higgs theory. In classifying the intrinsic topological orders, all the gauge fluxes are regarded as dynamical deconfined objects. Two (Abelian) topological phases are equivalent as long as they have the same quasiparticle braiding matrices (so-called $T$ and $S$ matrices), regardless of how the quasiparticles are labeled. This kind of equivalence relation is nothing but the $\mathbb{GL}(N,\mathbb{Z})$ equivalence for the $\vec{K}$ matrix.

To illustrate the difference, first we start from $G=\mathbb{Z}_n$. As disucussed in the main text, the $n$ different SPTs after being gauged result in the following $n$ gauge theories:
\begin{equation}
  \vec{K}=
  \begin{pmatrix}
	0 & n\\
	n & 2p
  \end{pmatrix}, p=0,1,\dots, n-1.
  \label{}
\end{equation}
One might wonder these gauge theories are all distinct. However, this is not generally true. In fact, for every odd $n\geq 5$, we have the following $\mathbb{GL}(2,\mathbb{Z})$ equivalence between $p=2$ and $p=\frac{n+1}{2}$:
\begin{equation}
  W^T\begin{pmatrix}
	0 & n\\
	n & 4
  \end{pmatrix}W=
  \begin{pmatrix}
	0 & n\\
	n & n+1
  \end{pmatrix},
  W=
  \begin{pmatrix}
	-2 & -1\\
	n & \frac{n+1}{2}
  \end{pmatrix}.
  \label{}
\end{equation}
This is just one example and there could be more ``collapse'' for general $n$.

Our second example is the gauge group
 $G=\mathbb{Z}_2\times\mathbb{Z}_2$. The dynamical gauged theory is given by the following $\mathbf{K}$ matrices:
\begin{equation}
  \mathbf{K}=
  \begin{pmatrix}
	\mathbf{0}_{2\times 2} & \mathbf{2}_{2\times 2}\\
	\mathbf{2}_{2\times 2} & \tilde{K}
  \end{pmatrix},
  \tilde{K}=
  \begin{pmatrix}
	k & l\\
	l & p
  \end{pmatrix}, k,p\in\{0,2\}, l\in \{0,1\}.
  \label{eqn:z2z2}
\end{equation}
The possible choices of $k, p$ and $l$ yields all the $H^3(\mathbb{Z}_2\times\mathbb{Z}_2, \mathbb{U}(1))=\mathbb{Z}_2^3$ cohomology classes.

 In fact, the $8$ classes listed in \eqref{eqn:z2z2} reduce to only $4$ under generic $\mathbb{SL}(4,\mathbb{Z})$ equivalence. In terms of the $\tilde{K}$ matrices, there are only the following four different intrinsic topological orders:
\begin{equation}
  \begin{gathered}
\begin{pmatrix}
	0 & 0\\
	0 & 0
  \end{pmatrix},  \begin{pmatrix}
	2 & 1\\
	1 & 2
  \end{pmatrix}
\\
  \begin{pmatrix}
	2 & 0\\
	0 & 0
  \end{pmatrix}\sim\begin{pmatrix}
	0 & 0\\
	0 & 2
  \end{pmatrix}
\sim\begin{pmatrix}
	2 & 0\\
	0 & 2
  \end{pmatrix},\\
\begin{pmatrix}
	0 & 1\\
	1 & 0
  \end{pmatrix}\sim
\begin{pmatrix}
	2 & 1\\
	1 & 0
  \end{pmatrix}\sim\begin{pmatrix}
	0 & 1\\
	1 & 2
  \end{pmatrix}.
\end{gathered}
  \label{}
\end{equation}
Here $\sim$ denotes the equivalence of the derived intrinsic topological orders.

In the response theory, we are not allowed to permute the gauge fluxes from different subgroups of the symmetry group since they correspond to different physical symmetries.

\section{Equivalence between fermionic and bosonic SPT phases with $\mathbb{Z}_{m}$ symmetry }

We demonstrate directly that when $m$ is odd all $\mathbb{Z}_m$ fermionic SPT phase are equivalent to bosonic ones. Let us consider the bulk Chern-Simons theory of $\mathbb{Z}_m$ fermionic SPT phase
\begin{equation}
  \mathbf{K}=
  \begin{pmatrix}
	1 & 0\\
	0 & -1
  \end{pmatrix}, \vec{q}_g=
  \begin{pmatrix}
	q_1\\
	q_2
  \end{pmatrix}.
  \label{}
\end{equation}
Here $g$ denotes the generator of the $\mathbb{Z}_m$ symmetry. We then add a trivial phase given by $\vec{K}=\sigma_z$ with trivial symmetry transformation on the edge bosons  $\vec{q}'_g=\begin{pmatrix} p \\ p\end{pmatrix}$ where $p\in \mathbb{Z}$. We pack the whole system into a $4\times 4$ $\vec{K}$ given by $\vec{K}=\sigma_z\otimes \mathbf{1}_{2\times 2}$.

 We then perform the following $\mathbb{SL}(4, \mathbb{Z})$ transformation
  \begin{equation}
	\vec{W}=
	\begin{pmatrix}
	  1 & 0 & 1 & 0\\
	  0 & 1 & 0 & 0\\
	  -1 & 0 & 0 & 1\\
	  1 & 0 & 1 & -1
	\end{pmatrix}.
	\label{}
  \end{equation}
  Under $\vec{W}$ the $\vec{K}$ matrix becomes $\vec{K}=\begin{pmatrix} \sigma_z & 0\\ 0 & \sigma_x\end{pmatrix}$. So the first two components are describing fermionic systems and the last two bosonic ones. We denote the symmetry vector for the collectively as $\tilde{\vec{q}}_g=(q_1,q_2, p, p)^T$. The edge modes in the new basis are denoted by
	\begin{equation}
	  \tilde{\bm{\phi}}=
	  \begin{pmatrix}
		\phi_1^f\\
		\phi_2^f\\
		\phi_1^b\\
		\phi_2^b
	  \end{pmatrix}
	  =
	  \begin{pmatrix}
		\phi_1-\phi_1'-\phi_2'\\
		\phi_2\\
		\phi_1'+\phi_2'\\
		\phi_1-\phi_2'
	  \end{pmatrix}
	  \label{}
	\end{equation}
	Under the $\mathbb{SL}(4, \mathbb{Z})$ transformation $\vec{W}$, the vector $\tilde{\vec{q}}_g \rightarrow \mathbf{W}^{-1}\tilde{\vec{q}}_g=(q_1-2p, q_2, 2p, q_1-p)^T$.
	
	If $q_1-q_2$ is even, we let $p=\frac{q_1-q_2}{2}$ and then $\tilde{\vec{q}}_g=(q_2, q_2, q_1-q_2, \frac{q_1+q_2}{2})^T$.
	Thus the fermionic SPT is equivalent to a bosonic one with $\mathbf{K}=\sigma_x, {\vec{q}}^b_g=(q_1-q_2,\frac{q_1+q_2}{2})^T$.

	If $q_1-q_2$ is odd, it seems that $p=\frac{q_1-q_2}{2}$, being a half integer, is not physical. Here the fermionic nature plays a crucial. This is most easily understood from the edge modes. The edge modes $\bm{\phi}'=(\phi_1', \phi_2')^T$ transforms under the $\mathbb{Z}_m$ symmetry as
	\begin{equation}
	  U_g\bm{\phi}'U_g^\dag= \bm{\phi}'+\frac{2\pi p}{m}
	  \begin{pmatrix}
		1\\
		1
	  \end{pmatrix}.
	  \label{}
	\end{equation}
	When $p$ is a half integer, we have
	\begin{equation}
	U_g^m\bm{\phi}'(U_g^\dag)^m=\bm{\phi}'+2\pi p\begin{pmatrix}
	  1\\ 1
	\end{pmatrix}
\equiv \bm{\phi}'+\pi
	\begin{pmatrix}
	  1\\ 1
	\end{pmatrix}
	,
	  \label{}
	\end{equation}
which is projectively the identity in a fermionic system.

	We now turn to the corresponding bosonic SPT. We notice that
	\begin{equation}
	  U_g^m \bm{\phi}^b(U_g^\dag)^m=\bm{\phi}^b+\pi \begin{pmatrix}
	  0\\
	  q_1+q_2
	\end{pmatrix}
	  \label{}
	\end{equation}
	which is not consistent with $U_g^m=1$ in the bosonic case. Again the identity transformation in a fermionic system can be realized projectively: 
	\begin{equation}
	  \begin{gathered}
	  \phi_1\rightarrow \phi_1+\pi\\
	  \phi_2\rightarrow\phi_2+\pi,
	\end{gathered}
	  \label{}
	\end{equation}
	 which means that 
	 \begin{equation}
	   \phi^b_2=\phi_1-\phi_2'\rightarrow \phi^b_2+\pi
	   \label{}
	 \end{equation}
	 is also an identity transformation. So we can freely add $\begin{pmatrix} 0\\ \pi\end{pmatrix}$ to $\bm{\phi}^b$ and as a result $U_g^m$ can differ by $\begin{pmatrix} 0 \\ m\pi\end{pmatrix}$, which makes the derived relation legitimate for odd $m$, but not for even $m$. We therefore prove that the fermionic SPT phases with $\mathbb{Z}_m$ are all equivalent to bosonic ones when $m$ is odd.

\end{document}